\documentclass[twocolumn,prl,showpacs,showkeys,amsmath,amssymb]{revtex4}
\usepackage{graphicx,amsmath,epsfig,latexsym,color}
\newcommand{\be}{\begin{equation}}
\newcommand{\ee}{\end{equation}}
\begin{document}
\title{Observation of atom wave phase shifts induced by van der Waals atom-surface interactions}
\author{John D.\ Perreault and Alexander D.\ Cronin}
\affiliation{University of Arizona, Tucson, Arizona 85721}
\date{\today}
\begin{abstract}
The development of nanotechnology and atom optics relies on
understanding how atoms behave and interact with their
environment. Isolated atoms can exhibit wave-like (coherent)
behaviour with a corresponding de Broglie wavelength and phase
which can be affected by nearby surfaces. Here an atom
interferometer is used to measure the phase shift of Na atom waves
induced by the walls of a 50 nm wide cavity.  To our knowledge
this is the first direct measurement of the de Broglie wave phase
shift caused by atom-surface interactions.  The magnitude of the
phase shift is in agreement with that predicted by quantum
electrodynamics for a non-retarded van der Waals interaction. This
experiment also demonstrates that atom-waves can retain their
coherence even when atom-surface distances are as small as 10 nm.
\end{abstract}
 \maketitle

The generally accepted picture of the electromagnetic vacuum
suggests that there is no such thing as empty space.  Quantum
electrodynamics tells us that even in the absence of any free
charges or radiation the space between atoms is actually permeated
by fluctuating electromagnetic fields.  An important physical
consequence of this view is that the fluctuating fields can
polarize atoms resulting in a long range attractive force between
electrically neutral matter:  the van der Waals (vdW) interaction
\cite{milo94}. This microscopic force is believed to be
responsible for the cohesion of nonpolar liquids, the latent heat
of many materials, and deviations from the ideal gas law.  The vdW
interaction can also affect individual atoms or groups of atoms
near a solid surface. For example, nearby surfaces can distort the
radial symmetry of carbon nanotubes \cite{ruof93} and deflect the
probes of atomic force microscopes \cite{gies03}. Atom-surface
interactions can also be a source of quantum decoherence or
uncontrolled phase shifts, which are an important considerations
when building practical atom interferometers on a chip
\cite{folm01}. For the case of an atom near a surface the vdW
potential takes the form $V(r)=-C_{3}r^{-3}$, where $C_{3}$
describes the strength of the interaction and r is the
atom-surface distance \cite{milo94}.

Previous experiments have shown how atom-surface interactions
affect the \emph{intensity} of atom waves transmitted through
cavities \cite{shih75,ande88,suke93}, diffracted from material
gratings \cite{gris99,bruh02,cron04,perr05,brez02}, and reflected
from surfaces \cite{ande86,berk89,shim01}. However, as we shall
see, none of these experiments provide a complete characterization
of how atom-surface interactions alter the \emph{phase} of atom
waves. In order to monitor the phase of an atom wave one must have
access to the wave function itself ($\psi$), not just the
probability density for atoms ($|\psi|^{2}$).  In this Letter an
atom interferometer is used to directly observe how atom surface
interactions affect the phase of atom waves.  This observation is
significant because it offers a new measurement technique for the
vdW potential and is of practical interest when designing atom
optics components on a chip \cite{henk99,folm02}.

When an atom wave propagates through a cavity it accumulates a
spatially varying phase due to its interaction with the cavity
walls\be
\begin{split}
\phi(\xi) \equiv \phi_{o} + \delta\phi(\xi) =
-\frac{lV(\xi)}{\hbar v},
\end{split}
\label{eq:phixi}\ee where $l$ is the interaction length, $V(\xi)$
is the atom-surface potential within the cavity, $\hbar$ is
Plank's constant, and $v$ is the particle velocity \cite{perr05}.
\begin{figure}
\scalebox{.65}{\includegraphics{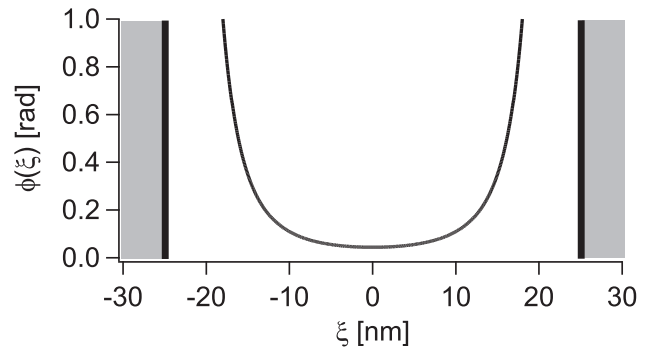}}
\caption{\label{fig:phivsxsi}Accumulated phase $\phi(\xi)$ of an
atom wave as a function of position due to a vdW interaction with
$C_{3}=$ 3 meV nm$^{3}$.  The atom wave has propagated through a
150 nm long cavity at a velocity of 2 km/s. The grey rectangles
indicate the location of the cavity walls which are 50 nm apart.
Notice how there is a non-zero constant phase offset
$\phi_{o}\sim$ 0.05 rad.}
\end{figure}
Equation \ref{eq:phixi} also separates the induced phase
$\phi(\xi)$ into constant $\phi_{o}$ and spatially varying
$\delta\phi(\xi)$ parts.  A plot of the phase $\phi(\xi)$ from
Eqn. \ref{eq:phixi} is shown in Fig. \ref{fig:phivsxsi} for the
cavity geometry and vdW interaction strength in our experiment. If
these cavities have a width $w$ and are oriented in an array with
spacing $d$, then the atom wave far away will have spatially
separated components (diffraction orders) with amplitudes\be
\begin{split}
\psi_{n}=A_{n}e^{i\Phi_{n}}=e^{i\phi_{o}}\int_{-w/2}^{w/2}e^{i\delta\phi(\xi)}e^{i2\pi\xi\frac{n}{d}}d\xi,
\end{split}
\label{eq:psi}\ee where $A_{n}$ and $\Phi_{n}$ are real numbers,
and $n$ is the diffraction order number \cite{perr05}. Experiments
which measure the intensity of atom waves (e.g. atom wave
diffraction) are only sensitive to $|\psi_{n}|^{2}=|A_{n}|^{2}$
which is in part influenced by $\delta\phi(\xi)$. However, it is
clear from Eqn. \ref{eq:psi} that $|\psi_{n}|^{2}$ reveals no
information about $\phi_{o}$ or $\Phi_{n}$.  We have determined
$A_{0}$ and $\Phi_{0}$ by placing this array of cavities (grating)
in one arm of an atom interferometer. This new technique is
sensitive to the \emph{entire} phase shift $\phi(\xi)$ induced by
an atom-surface interaction, including the constant offset
$\phi_{o}$.

\begin{figure}
\scalebox{.5}{\includegraphics{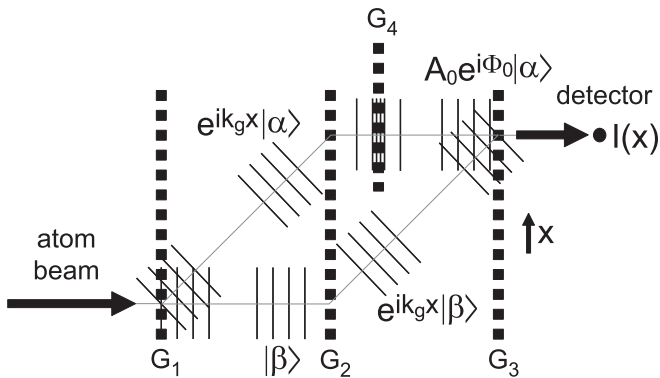}}
\caption{\label{fig:setupfig}Experimental setup for vdW induced
phase measurement.  A Mach-Zhender atom interferometer with paths
$\alpha$ and $\beta$ is formed using the zeroeth and first order
diffracted beams of gratings $G_{1}$ and $G_{2}$ which have a
period of 100 nm.  The atom wave interference pattern is read out
using grating $G_{3}$ as an amplitude mask.  The phase fronts
(groups of parallel lines) passing through grating $G_{4}$ are
compressed due to the attractive vdW interaction, resulting in a
phase shift $\Phi_{o}$ of beam $|\alpha\rangle$ relative to
$|\beta\rangle$. This causes the interference pattern I(x) to
shift in space at the plane defined by $G_{3}$.}
\end{figure}

The experimental setup for using an atom interferometer to measure
the phase shift $\Phi_{0}$ induced by atom-surface interactions is
shown in Fig. \ref{fig:setupfig}.
A beam of Na atoms travelling at roughly 2 km/s is generated from
an oven and a position state of the atom wave is selected by two
10 $\mu$m collimation slits spaced 1 m apart.  A Mach-Zehnder type
interferometer is formed using the zeroeth and first order
diffracted beams from three 100 nm period silicon nitride gratings
\cite{sava96}.  The grating $G_{1}$ creates a superposition of
position states $|\alpha\rangle$ and $|\beta\rangle$ which
propagate along separated paths $\alpha$ and $\beta$ respectively.
The states are then recombined using grating $G_{2}$ and form a
spatial interference pattern at the plane of $G_{3}$. The third
grating $G_{3}$ is used as an amplitude mask to determine the
phase and contrast of the interference pattern.  A co-propagating
laser interferometer (not shown in Fig. \ref{fig:setupfig}) was
used to compensate for mechanical vibrations of
$G_{1},G_{2},G_{3}$.

When grating $G_{4}$ is inserted into the interferometer path
$\alpha$, the interference pattern I(x) shifts in space along the
positive x-direction.  This can be understood by recalling de
Broglie's relation $\lambda_{dB}=h/p$ \cite{berm97,meys01}.  The
atoms are sped up by the attractive vdW interaction between the Na
atoms and the walls of grating $G_{4}$. This causes $\lambda_{dB}$
to be smaller in the region of $G_{4}$, compressing the atom wave
phase fronts and retarding the phase of beam $|\alpha\rangle$ as
it propagates along path $\alpha$.  One could also say that
$G_{4}$ effectively increases the optical path length of path
$\alpha$.  At $G_{3}$ the beams $|\alpha\rangle$ and
$|\beta\rangle$ then have a relative phase between them leading to
the state\be
\begin{split}
|\chi\rangle = A_{0}e^{i\Phi_{0}}|\alpha\rangle +
e^{ik_{g}x}|\beta\rangle,
\end{split}
\label{eq:chi}\ee where $k_{g}=2\pi/d$.  The diffraction amplitude
$A_{0}$ reflects the fact that beam $|\alpha\rangle$ is also
attenuated by $G_{4}$.  The state $|\chi\rangle$ in Eqn.
\ref{eq:chi} then leads to an interference pattern that is shifted
in space by an amount that depends on $\Phi_{0}$\be
\begin{split}
I(x) = \langle\chi|\chi\rangle \propto 1 + C\cos(k_{g}x-\Phi_{0}),
\end{split}
\label{eq:interference}\ee where $C$ is the contrast of the
interference pattern.  Inserting $G_{4}$ into path $\beta$ will
result in the same form of the interference pattern in Eqn.
\ref{eq:interference}, but with a phase shift of the opposite sign
(i.e. $\Phi_{0}\rightarrow-\Phi_{0}$).

Grating $G_{4}$ is an array of cavities 50 nm wide and 150 nm long
which cause a potential well for the Na atoms due to the vdW
interaction.  The atoms transmitted through $G_{4}$ must pass
within 25 nm of the cavity walls since the open slots of the
grating are 50 nm wide.  At this atom-surface distance the depth
of the potential well is about $4\times10^{-7}$ eV.  Therefore, as
the atoms enter the grating they are accelerated by the vdW
interaction energy from 2000 m/s to 2000.001 m/s and decelerated
back to 2000 m/s as they leave the grating.  This small change in
velocity is enough to cause about a 0.3 rad phase shift which
corresponds to a 5 nm displacement of the interference pattern. It
is quite remarkable to note that the acceleration and deceleration
happens over a time period of 75 ps implying that the atoms
experience an acceleration of at least $10^{6}$ g's while passing
through the grating. This indicates that the vdW interaction is
one of the most important forces at the nanometer length scale.

\begin{figure}
\scalebox{.45}{\includegraphics{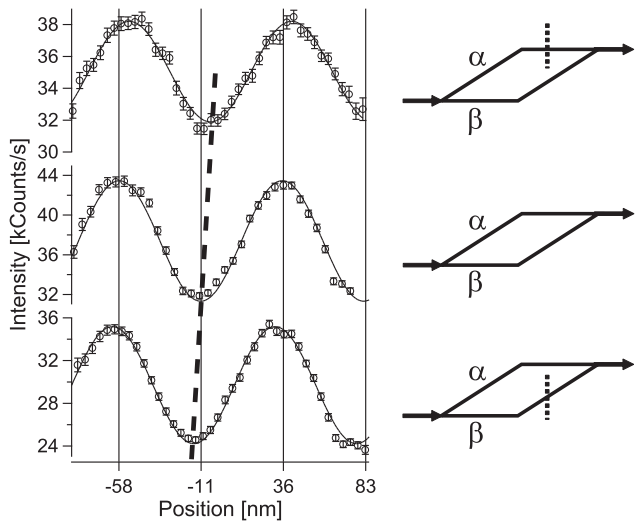}}
\caption{\label{fig:rawdata}Interference pattern observed when the
grating $G_{4}$ is inserted into path $\alpha$ or $\beta$ of the
atom interferometer.  Each interference pattern represents 5
seconds of data.  The intensity error bars are arrived at by
assuming Poisson statistics for the number of detected atoms.  The
dashed line on the plots is a visual aid to help illustrate the
measured phase shift of 0.3 radians.  Notice how the phase shift
induced by placing $G_{4}$ in path $\alpha$ or $\beta$ has
opposite sign.  The sign of the phase shift is also consistent
with the atom experiencing an attractive potential as it passes
through $G_{4}$.}
\end{figure}

The experiment consists of measuring shifts in the position of the
interference pattern I(x) when $G_{4}$ is moved in and out of the
interferometer paths.  The interference data is shown in Fig.
\ref{fig:rawdata}.
When $G_{4}$ is placed in path $\alpha$ the fringes shift in the
positive x-direction, whereas placing $G_{4}$ in path $\beta$
causes a shift in the negative x-direction. Therefore the absolute
sign of the phase shift is consistent with an attractive force
between the Na atoms and the walls of grating $G_{4}$.  It is also
observed that although the Na atoms are passing within 25 nm of
the grating the atom waves retain their wave like behaviour
(coherence), as evident by the non-zero contrast of the
interference fringes.

The atom interferometer had a linear background phase drift of
approximately $2\pi$ rad/hr and non-linear excursions of $\sim$1
rad over a period of 10 min, which were attributed to thermally
induced mechanical drift of the interferometer gratings
$G_{1},G_{2},G_{3}$ and phase instability of the vibration
compensating laser interferometer. The data were taken by
alternating between test ($G_{4}$ in path $\alpha$ or $\beta$) and
control ($G_{4}$ out of the interferometer) conditions with a
period of 50 seconds, so that the background phase drift was
nearly linear between data collection cycles.  A fifth order
polynomial was fit to the phase time series for the control cases
and then subtracted from the test and control data. All of the
interference data was corrected in this way.

Grating $G_{4}$ had to be prepared so that it was possible to
obscure the test arm of the interferometer while leaving the
reference arm unaffected. The grating is surrounded by a silicon
frame, making it necessary to perforate $G_{4}$.  The grating bars
themselves are stabilized by 1 $\mu$m period support bars running
along the direction of $\textbf{k}_{\textbf{g}}$. The grating
naturally fractured along these support structures after applying
pressure with a drawn glass capillary tube.  Using this
preparation technique $G_{4}$ had a transition from intact grating
to gap over a distance of about 3 $\mu$m, easily fitting inside
our interferometer which has a path separation of about 50 $\mu$m
for atoms travelling at 2 km/s.

Due to the preparation technique, $G_{4}$ was inserted into the
test arm with $\textbf{k}_{\textbf{g}}$ orthogonal to the plane of
the interferometer.  This causes diffraction of the test arm out
of the plane of the interferometer, in addition to the zeroeth
order. However, the diffracted beams have an additional path
length of approximately 2 nm due to geometry.  Since our atom beam
source has a coherence length of
$\lambda_{dB}\frac{v}{\sigma_{v}}=$ 0.1 nm, the interference
caused by the diffracted beams will have negligible contrast.
Therefore, the zeroeth order of $G_{4}$ will be the only
significant contribution to the interference signal.

\begin{figure}
\scalebox{.7}{\includegraphics{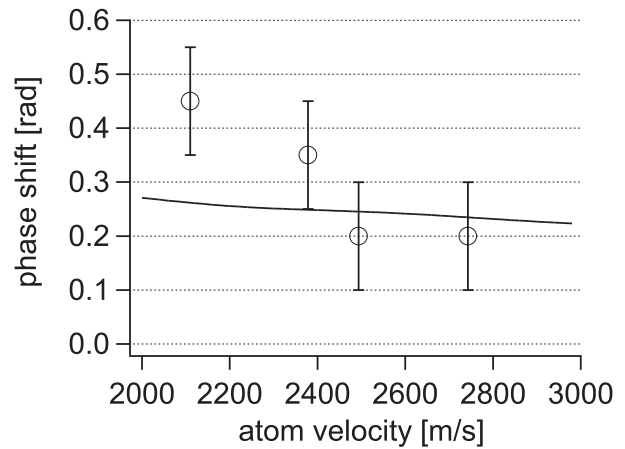}}
\caption{\label{fig:phivsv}Phase shift $\Phi_{0}$ induced by
grating $G_{4}$ for various atom beam velocities.  The phase shift
data has been corrected for systematic offsets ($\sim$30 \%)
caused by the interference of other diffraction orders in the atom
interferometer and the error bars reflect the uncertainty in the
systematic parameters. The solid line is a prediction of the
induced phase shift for vdW coefficient $C_{3}=$ 3 meV nm$^{3}$,
grating thickness 150 nm, and grating open fraction 0.5.  The data
agrees in magnitude with the prediction and reproduces the slight
trend of decreasing phase shift with increasing velocity.}
\end{figure}

In principle the amount of phase shift $\Phi_{0}$ induced by the
vdW interaction should depend on how long the atom spends near the
surface of the grating bars.  Therefore the observed phase shift
produced by placing $G_{4}$ in one of the interferometer paths
should depend on the atom beam velocity in the way described by
Eqs. \ref{eq:phixi} and \ref{eq:psi}. To test this prediction the
experiment illustrated in Fig. \ref{fig:rawdata} was repeated for
several different atom beam velocities and the data are shown in
Fig. \ref{fig:phivsv}.
Systematic phase offsets of ($\sim$30 \%) caused by the detected
interference of additional diffraction orders generated by
$G_{1},G_{2},G_{3}$ in the atom interferometer (not shown in Fig.
\ref{fig:setupfig}) have been corrected for in Fig.
\ref{fig:phivsv}.  The uncertainty of the phase measurements in
Fig. \ref{fig:phivsv} is given by the uncertainty in the
systematic parameters. The measured phase shift compares well to a
prediction of the phase shift $\Phi_{0}$ for the zeroeth order of
grating $G_{4}$ which includes the vdW interaction.  The value of
$C_{3}=$ 3 meV nm$^{3}$ used to generate the theoretical
prediction in Fig. \ref{fig:phivsv} is consistent with previous
measurements based on diffraction experiments \cite{perr05}.  It
is important to note that if there was no interaction between the
atom and the grating there would be zero observed phase shift.

The confirmation of atom-surface induced phase shifts presented
here can be extrapolated to the case of atoms guided on a chip.
Atoms travelling at 1 m/s over a distance of 1 cm will have an
interaction time of 0.01 seconds.  According to Eqn.
\ref{eq:phixi}, if these atoms are 0.1 $\mu$m from the surface
they will acquire a phase shift of $5\times 10^{4}$ radians due to
the vdW interaction.  Similarly, if the atoms are 0.5 $\mu$m from
the surface they will have a phase shift of $4\times 10^{2}$
radians. Therefore, a cloud of atoms 0.1 $\mu$m from a surface
will have a rapidly varying phase profile which could severely
reduce the contrast of an interference signal.  At some
atom-surface distance the vdW interaction will significantly alter
atom-chip trapping potentials, resulting in loss of trapped atoms.
Atom-chip magnetic traps are harmonic near their center and can
have a trap frequency of $\omega = 2\pi\times 200$ kHz
\cite{folm02}. Given the vdW interaction we have observed, such a
magnetic trap would have no bound states for Na atoms if its
center was closer than 220 nm from a surface. Therefore, the vdW
interaction places a limit on the spatial scale of atom
interferometers built on a chip because bringing the atoms too
close to a surface can result in poor contrast and atom intensity.

In conclusion the affect of atom-surface interactions on the phase
of a Na atom wave has been observed directly for the first time.
When the atom wave passes within 25 nm of a surface for 75 ps it
accumulates a phase shift of $\Phi_{0}\approx$ 0.3 rad consistent
with an attractive vdW interaction. The slight velocity dependence
of this interaction has also been confirmed. This experiment has
also demonstrated the non-obvious result that atom waves can
retain their coherence when passing within 25 nm of a surface.  In
the future one could use this experiment to make a more precise
measurement of $C_{3}$ at the 10 \% level if the interference of
unwanted diffraction orders are eliminated and the window size $w$
of $G_{4}$ is determined with a precision of \mbox{3 \%}. This
level of precision in measuring $w$ is possible with existing
scanning electron microscopes.

This research was supported by grants from Research Corporation
and the National Science Foundation.


\end{document}